\documentclass[aps,prb,twocolumn,amsmath,amssymb,showpacs,superscriptaddress]{revtex4}

\usepackage{bm}
\usepackage{graphicx}
\usepackage{color}

\begin{document}


\title{Linear magnetoresistance on the topological surface}

\author{C. M. Wang}
\email[]{cmwangsjtu@gmail.com}
\affiliation{School of Physics and
Electrical Engineering, Anyang Normal University, Anyang 455000,
China}
\author{X. L. Lei}
\affiliation{Department of Physics, Shanghai Jiaotong University,
1954 Huashan Road, Shanghai 200030, China}

\date{\today}

\begin{abstract}
A positive, non-saturating and dominantly linear magnetoresistance
is demonstrated to occur in the surface state of a topological
insulator having a wavevector-linear energy dispersion together
with a finite positive Zeeman energy splitting. This linear magnetoresistance shows up
within quite wide magnetic-field range in a spatially homogenous system of high carrier density
and low mobility in which the conduction electrons are in extended states and spread over
many smeared Landau levels, and is robust against increasing temperature,
in agreement with recent experimental findings in Bi$_2$Se$_3$ nanoribbons.

\end{abstract}


\maketitle

\section{introduction}
It is well known that the classical magnetoresistance (MR) in metals
or semiconductors with a closed free electron Fermi surface increases
quadratically with increasing magnetic field $B$ for $\mu B\ll 1$ and saturates when
$\mu B > 1$. Here $\mu$ is the zero-magnetic-field mobility. Hence,
the extraordinarily high and linear MR (LMR), which breaks this
familiar rule, has been gaining much attention as soon as its
discovery. In the past decade, this unexpected LMR has been reported
in silver chalcogenide,\cite{xu1997large} indium
antimonide,\cite{hu2008classical} silicon,\cite{delmo2009large}
MnAs-GaAs composite material,\cite{Johnson2010} and
graphene.\cite{friedman2010quantum}

Kapitza's linear law\cite{Kapitza1929} indicates that the metal shows
a magnetoresistance linear in perpendicular magnetic field when it
has an open Fermi surface and a mean free path longer than the
electronic Larmor radius. Recently, another two models, irrespective
of the open Fermi surface, have been constructed to provide possible
mechanisms for the LMR phenomenon. Abrikosov suggested a quantum-limit origin of
LMR for the homogenous system with a gapless linear energy
spectrum.\cite{Abrikosov1998,abrikosov2000quantum} His model
requires that Landau levels
are well formed and the carrier concentration is small that all electrons
occupy only the lowest Landau band. Alternatively, Parish and Littlewood developed a
classical model without involving linear spectrum.\cite{Parish2003}
Ignoring the concrete microscopic
mechanism, they attributed this unusual MR to the mobility
fluctuations in a strongly inhomogenous system.

Topological
insulators\cite{Kane2005,hasan2010colloquium,qi2010topological}
(TIs) are novel materials with a full energy gap in bulk, while
there are gapless surface states. Due to its unique band structure
with only one helical Dirac cone and linear energy
dispersion,\cite{xia2009observation,zhang2009natphys,CXLiu2010} the
surface states of the TI Bi$_2$Se$_3$ become an excellent platform
for the study of quantum-limit LMR. The recent experiment in this
flat surface system, however, reported that a large positive MR,
which becomes very linear above a characteristic field of
$1$$\sim$$2$\,T, was observed even in an opposite situation where
the carrier sheet density is high that electrons occupy more than
one Landau levels.\cite{Tang2011} Moreover, they found that raising
temperature to room temperature almost has no influence on the
observed LMR. It is striking that this observation is in conflict
with Abrikosov's model and also with the classical Parish-Littlewood
model.
 So far a reliable theoretical scheme capable of explaining this novel experiment has still been lacking.

In this paper, we generalize the balance-equation
approach\cite{lei1985gsf} to a system modeling the surface states of
a three-dimensional TI to investigate the two-dimensional
magnetotransport in it. We find that a  positive, nonsaturating and
dominantly linear magnetoresistance can appear within quite wide
magnetic-field range in the TI surface state having a positive and
finite effective g-factor. This linear magnetoresistance shows up in
the system of high carrier concentration and low mobility when
electrons are in extended states and spread over many smeared Landau
levels, and persists up to room temperature, providing a possible
mechanism for the recently observed linear magnetoresistance in
topological insulator Bi$_2$Se$_3$ nanoribbons.\cite{Tang2011}

\section{Balance-Equation Formulation for magnetoresistivity}
We consider the surface state of a Bi$_2$Se$_3$-type large bulk gap
TI in the $x$-$y$ plane under the influence of a uniform magnetic
field $\bm B$ applied along the $z$ direction.\cite{CXLiu2010}
Following the experimental observation,\cite{Tang2011} we assume
that the Fermi energy locates in the gap of the bulk band and above
the Dirac point, i.e. the surface carriers are electrons. Further,
the separations of the Fermi energy from the bottom of bulk band and
Dirac point are much larger than the highest temperature ($300\,{\rm
K}$) considered in this work. Hence, the contribution from the bulk
band to the magnetotransport is negligible. These electrons,
scattered by randomly distributed impurities and by phonons, are
driven by a uniform in-plane electric field $\bm E=(E_x,E_y)$ in the
topological surface. The Hamiltonian of this many-electron and
phonon system consists of an electron part $\mathcal H_{\rm e}$, a
phonon part $\mathcal H_{\rm ph}$, and electron-impurity and
electron-phonon interactions $\mathcal H_{\rm ei}$ and $\mathcal
H_{\rm ep}$:
\begin{equation}\label{}
\mathcal H=\mathcal H_{\rm e}+\mathcal H_{\rm ei}+\mathcal H_{\rm
ep}+\mathcal H_{\rm ph}.
\end{equation}
Here, the electron Hamiltonian is taken in the form
\begin{equation}\label{helectron}
\mathcal H_{\rm e}=\sum_j\left[v_{\rm
F}(\pi_{j}^x\sigma_j^y-\pi_{j}^y\sigma_j^x) +\frac{1}{2}g_z\mu_{\rm
B}B \sigma_{j}^z+ e\bm r_j\cdot\bm E \right] ,
\end{equation}
in which ${\bm \pi}_j\equiv \bm p_j+e\bm A(\bm r_j)=(\pi_j^x, \pi_j^y)$, ${\bm r}_j=(x_{j},y_{j})$,
${\bm p}_j=(p_{jx},p_{jy})$ and ${\bm \sigma}_j=(\sigma_{j}^x,\sigma_{j}^y,\sigma_{j}^z)$,
 stand, respectively, for the canonical momentum, coordinate, momentum and spin operators
of the $j$th electron having charge $-e$, $\bm A(\bm r)=(-By,0)$ is the vector potential
of the perpendicular magnetic field $\bm B=B\hat{z}$ in the Landau gauge,
$v_{\rm F}$ is the Fermi velocity, $g_z$ is the effective g-factor of the surface electron, and
$\mu_{\rm B}=e/2m_0$ is the Bohr magneton with $m_0$ the free electron mass.
The sum index $j$ in Eq.\,(\ref{helectron}) goes over all electrons of total number $N$ in the surface state of unit area.

In the frame work of balance equation approach,\cite{lei1985tdb,lei1985gsf,cai1985}
the two-dimensional center-of-mass (c.m.) momentum and coordinate
$\bm P=\sum_j\bm p_j$ and $\bm R=N^{-1}\sum_j\bm r_j$, and the
relative-electron momenta and coordinates $\bm p'_j=\bm p_j-\bm P/N$
and $\bm r_j'=\bm r_j-\bm R$ are introduced to write the
Hamiltonian $\mathcal H_{\rm e}$ into the sum of a single-particle c.m. part
$\mathcal H_{\rm cm}$ and a many-particle relative-electron part $\mathcal H_{\rm
er}$: $\mathcal H_{\rm e}=\mathcal H_{\rm cm}+\mathcal H_{\rm er}$, with
\begin{align}\label{}
 \mathcal H_{\rm cm}&=v_{\rm F}(\varPi_x\sigma_{\rm c}^y-\varPi_y\sigma_{\rm c}^x)+Ne{\bm E}\cdot \bm R ,\\
\mathcal H_{\rm er}&=\sum_j\left[v_{\rm F}(\pi_j'^{x}\sigma_j^y-\pi_j'^{y}\sigma_j^x)+\frac{1}{2}g_z\mu_{\rm B}B\sigma_{j}^z\right].
\end{align}
In this, ${\bm \varPi}\equiv \bm P+Ne\bm A(\bm R)=(\varPi_x,
\varPi_y)$ is the canonical momentum of the center-of-mass and ${\bm
\pi}_j' \equiv \bm p_j'+e\bm A(\bm r_j')=(\pi_j'^x, \pi_j'^y)$ is
the canonical momentum for the $j$th relative electron. Here we have
also introduced c.m. spin operators ${\sigma}_{\rm c}^x \equiv
N^{-1}\sum_j \sigma_j^x$ and ${\sigma}_{\rm c}^y \equiv N^{-1}\sum_j
\sigma_j^y$. The commutation relations between the c.m. spin
operators $\sigma_{\rm c}^x$ and $\sigma_{\rm c}^y$ and the spin
operators $\sigma_j^x$, $\sigma_j^y$ and $\sigma_j^z$ of the $j$th
electron  are of order of $1/N$: $[\sigma_{j}^{\beta_1}, \sigma_{\rm
c}^{\beta_2}]= N^{-1}2\,{\rm
i}\,\varepsilon_{\beta_1\beta_2\beta_3}\sigma_j^{\beta_3}$ with
$\beta_1,\beta_2,\beta_3=(x,y,z)$. Therefore, for a macroscopic
large $N$ system, the c.m. part $\mathcal H_{\rm cm}$ actually
commutes with the relative-electron part $\mathcal H_{\rm er}$ in
the Hamiltonian, i.e. the c.m. motion and the relative motion of
electrons are truly separated from each other. The couplings between
the two emerge only through the electron--impurity and
electron--phonon interactions. Furthermore, the electric field ${\bm
E}$ shows up only in $\mathcal H_{\rm cm}$. And, in view of
$[r'_{i\alpha},p'_{j\beta}]={\rm i}\delta_{\alpha
\beta}(\delta_{ij}-1/N)\simeq {\rm
i}\delta_{\alpha\beta}\delta_{ij}$, i.e. the relative-electron
momenta and coordinates can be treated as canonical conjugate
variables, the relative-motion part $\mathcal H_{\rm er}$ is just
the Hamiltonian of $N$ electrons in the surface state of TI in the
magnetic field without the presence of the electric field.

In terms of the c.m. coordinate ${\bm R}$ and the relative electron
density operator $\rho_{\bm q}= \sum_j {\rm e}^{{\rm i}\,{\bm
q}\cdot{\bm r}'_j}$, the electron--impurity and electron--phonon
interactions can be written as\cite{lei1985tdb,cai1985}
\begin{align}
\mathcal H_{\rm ei}=&\sum_{{\bm q}, a}U({\bm q})\,
{\rm e}^{{\rm i}\,{\bm q}\cdot \left({\bm R}-{\bm
r}_{a}\right)}\rho_{{\bm q}},\label{1-14}\\
\mathcal H_{\rm ep}=&\sum_{{\bm Q},\lambda}
M({\bm Q},\lambda)\,
\phi_{{\bm Q}\lambda}{\rm e}^{{\rm i}\,{\bm q}\cdot{\bm R}}\rho_{{\bm q}}.
\end{align}
Here $U({\pmb q})$  and $M({\pmb Q},\lambda)$ are respectively the impurity potential
(an impurity at randomly distributed position ${\bm r}_a$) and electron--phonon
coupling matrix element in the plane-wave
representation, and $\phi_{{\bm Q}\lambda}\equiv b_{{\bm Q}\lambda}^{}+b_{-{\bm Q}\lambda}^{\dagger}$ with $b_{{\pmb Q}\lambda}^{\dagger}$
 and $b_{{\bm Q}\lambda}^{}$
 being the creation and annihilation operators for a phonon of wavevector
${\bm Q}=({\bm q}, q_z)$ in branch
$\lambda$ having frequency
${\it \Omega}_{{\bm Q}\lambda}$.

The c.m. velocity (operator) ${\bm V}$ is the time variation of its
coordinate: ${\bm V}=\dot{\bm R}=-{\rm i}[{\bm R}, \mathcal H]=
v_{\rm F}(\sigma_{\rm c}^y\, \hat{i}-\sigma_{\rm c}^x\, \hat{j})$.
To derive a force-balance equation for steady state transport we
consider the Heisenberg equation for the rate of change of the c.m.
canonical momentum $\bm \varPi$:
\begin{equation}\label{dotpi}
\dot{\bm \varPi}=-{\rm i}\,[{\bm \varPi},\mathcal H]= - N e({\bm
V}\times {\bm B})- N e{\bm E}+{\bm {F}}_{\rm i}+{\bm {F}}_{\rm p},
\end{equation}
in which the frictional forces ${\bm {F}}_{\rm i}$ and ${\bm {F}}_{\rm p}$ share the same
expressions as given in Ref.\,\onlinecite{cai1985}.

The statistical average of the operator equation \eqref{dotpi} can
be determined to linear order in the electron--impurity and
electron--phonon interactions $\mathcal H_{\rm ei}$ and $\mathcal
H_{\rm ep}$ with the initial density matrix $\hat
\rho_0=Z^{-1}e^{-(\mathcal H_{\rm ph}+\mathcal H_{\rm er})/T}$ at
temperature $T$ when the in-plane electric field ${\bm E}$ is not
strong. For steady-transport states we have $\langle \dot{\bm
\varPi}\rangle=0$, leading to a force-balance equation of the form
\begin{equation}\label{}
0=-Ne\bm v\times\bm B-Ne\bm E+\bm f_{\rm i}+\bm f_{\rm p}.\label{forceEq}\\
\end{equation}
Here ${\bm v}=\langle {\bm V} \rangle$, the statistically averaged velocity of the moving center-of-mass,
is identified as the average rate of change of its position,
i.e. the drift velocity of the electron system driven by the electric field ${\bm E}$,
and $\bm f_{\rm i}$ and $\bm f_{\rm p}$ are frictional forces experienced  by the center-of-mass
 due to impurity and phonon scatterings:
\begin{align}\label{}
\bm f_{\rm i}=&\sum_{\bm q}\left|U(\bm q)\right|^2\bm q\varPi_2(\bm q,\omega_0),\\
\bm f_{\rm p}=&\sum_{\bm Q,\lambda}\left|M(\bm
Q,\lambda)\right|^2\bm q\varPi_2(\bm q,{\it \Omega}_{\bm
Q\lambda}+\omega_0)\nonumber\\
&\hspace{0.4cm}\times\left[n\Big(\frac{{\it \Omega}_{\bm
Q\lambda}}{T}\Big)-n\Big(\frac{{\it \Omega}_{\bm
Q\lambda}+\omega_0}{T}\Big)\right],\label{fp}
\end{align}
in which  $n(x)=(e^x-1)^{-1}$ is the Bose distribution function,
$\omega_0\equiv {\bm q}\cdot {\bm v}$, and $\varPi_2(\bm q,\omega)$
stands for the imaginary part of the Fourier spectrum of the
relative-electron density correlation function defined by
\begin{equation}
\varPi({\bm q}, t-t^{\prime})=-{\rm i\,}\theta(t-t^{\prime})
\big\langle\big[\rho_{{\bm q}}(t),\, \rho_{-{\bm
q}}(t^{\prime})\big]\big\rangle_{0},
\end{equation}
where $\rho_{{\bm q}}(t)={\rm e}^{{\rm i\,}\mathcal H_{\rm
er}t}\rho_{{\bm q}}\,{\rm e}^{-{\rm i\,}\mathcal H_{\rm er}t}$
and $\langle...\rangle_{0}$ denotes the statistical averaging over the initial density matrix $\hat \rho_0$.\cite{lei1985gsf}

The force-balance equation \eqref{forceEq} describes the steady-state two-dimensional
 magnetotransport in the surface state of a TI.
Note that the frictional forces $\bm f_{\rm i}$ and  $\bm f_{\rm p}$
are in the opposite direction of the drift velocity $\bm v$ and
their magnitudes are functions of $v=|{\bm v}|$ only. With the drift
velocity $\bm v=(v,0)$ in the $x$ direction, the force-balance
equation Eq.\,\eqref{forceEq} yields a transverse resistivity
$R_{xy}=-E_y/(Nev)=-B/(Ne)$, and a longitudinal resistivity
$R_{xx}=-E_x/(Nev)=-(f_{\rm i}+f_{\rm p})/(N^2e^2v)$. The linear one
is in the form
\begin{align}\label{}
R_{xx}=&-\frac{1}{N^2e^2}\sum_{\bm q}|U(\bm q)|^2q_x^2
\frac{\partial}{\partial \omega}\varPi_2(\bm
q,\omega)\big|_{\omega=0}\nonumber\\&-\frac{1}{2TN^2e^2}\sum_{\bm
Q,\lambda}|M(\bm Q,\lambda)|^2q_x^2
\varPi_2(\bm q,{\it \Omega}_{\bm Q\lambda})\nonumber\\
&\hspace{3.0cm}\times{\rm csch}^2\Big(\frac{{\it \Omega}_{\bm Q\lambda}}{2T}\Big).
\end{align}

\section{Density correlation function in the Landau representation}

For calculating the electron density correlation function
$\varPi_2(\bm q, \omega)$ we proceed  in the Landau
representation.\cite{cai1985,Ting1977} The Landau levels of the
single-particle Hamiltonian $h=v_{\rm
F}(\pi^x\sigma^y-\pi^y\sigma^x)+\frac{1}{2}g_z\mu_{\rm B}B\sigma^z$
of the relative-electron system in the absence of electric field are
composed of a positive ``$+$'' and a negative ``$-$''
branch\cite{McClure1956,Zheng2002,ZGWang2010,Zarea2005,Taskin2011}
\begin{equation}\label{LLlevel}
\varepsilon_{n}^{\pm}=\pm \sqrt{2n \varepsilon_s^2+\delta_z^2}=\pm \varepsilon_{n}\,\,\, (n=1,2,....)
\end{equation}
with $\varepsilon_s=v_F\sqrt{eB}$ and $\delta_z=-\frac{1}{2}g_z\mu_{\rm B} B$, and a zero ($n=0$) level
\begin{equation}
\varepsilon_{0}=\delta_z=-\frac{1}{2}g_z\mu_{\rm B}B.
\end{equation}
The corresponding Landau
wave functions are
\begin{equation}\label{}
\Psi_{n,k_x}^{+}(\bm r)=\frac{1}{\sqrt{\mathcal R_n}}e^{ik_xx}\left(
                                                    \begin{array}{c}
                                                      {\rm i}\mathcal P_{n}\phi_{{n-1},k_x}(y) \\
                                                      \phi_{n, k_x}(y) \\
                                                    \end{array}
                                                  \right)
\end{equation}
and
\begin{equation}\label{}
\Psi_{n,k_x}^{-}(\bm r)=\frac{1}{\sqrt{\mathcal R_n}}e^{ik_xx}\left(
                                                    \begin{array}{c}
                                                      \phi_{{n-1},k_x}(y) \\
                                                      {\rm i}\mathcal P_{n}\phi_{n, k_x}(y) \\
                                                    \end{array}
                                                  \right)
\end{equation}
for $n=1,2,...$; and
 \begin{equation}\label{}
\Psi_{0,k_x}(\bm r)= e^{{\rm i}k_xx}\left(
                                                    \begin{array}{c}
                                                                   0 \\
                                                      \phi_{0,k_x}(y) \\
                                                    \end{array}
                                                  \right)
\end{equation}
 for $n=0$.
Here $k_x$ is the wavevector of the
system along $x$ direction; $\mathcal R_n=1+\mathcal
P_n^2$ with $\mathcal
P_n=\sqrt{2n}\varepsilon_s/(\delta_z+\sqrt{2n\varepsilon_s^2+\delta_z^2})$;
and $\phi_{n,k_x}(y)=D_n\exp(-\gamma^2/2)H_n(\gamma)$ is the harmonic oscillator eigenfunction
with $H_n(x)$ being the Hermite polynomial, $\gamma\equiv (y-y_c)/l_{\rm B}=\sqrt{eB}(y-k_x l_{\rm B}^2)$,
and $D_n=1/(2^nn!)^{1/2}(eB/\pi)^{1/4}$.

Each Landau level contains $n_{\rm B}=eB/2\pi=1/(2\pi l_{\rm B}^{2})$ electron states
for system of unit surface area.
 The positive branch $\varepsilon_{n}^{+}=\varepsilon_{n}$ and the $n=0$ level $\varepsilon_{0}$
 of the above energy spectra
 are indeed quite close to those of the surface states in the bulk gap
 of Bi$_2$Se$_3$-family materials derived from microscopic band calculation.\cite{CXLiu2010}

The Landau levels are broadened due to impurity, phonon and
electron-electron scatterings. We model the imaginary part of the
retarded Green's function, or the density-of-states, of the
broadened Landau level $n$ (written for ``+''-branch and $n=0$
levels), using a Gaussian-type form:\cite{Ando1982}
\begin{equation}\label{}
{\rm Im}G_n(\epsilon)=-\frac{\sqrt{2\pi}}
\varGamma\exp\left[-\frac{2(\epsilon-\varepsilon_n)^2} {
\varGamma^2}\right],
\end{equation}
with a half-width ${\it \Gamma}$ of the form:\cite{Zheng2002}
$\varGamma=\left[{2\omega_c}/({\pi\tau_s})\right]^{1/2}$. Here
$\tau_s$ is the single-particle lifetime and $\omega_c=eBv_{\rm
F}^2/\varepsilon_{\rm F}^0$ is the cyclotron frequency of
linear-energy-dispersion system with $\varepsilon_{\rm F}^0=2v_{\rm
F}\sqrt{\pi N}$ being the zero-temperature Fermi level. Using a
semi-empirical parameter $\alpha$ to relate $\tau_s$ with the
transport scattering time $\tau_{\rm tr}=4\alpha\tau_{s}$, and
expressing $\tau_{\rm tr}$ with the zero-field mobility $\mu$ at
finite temperature,\cite{Dimitrie}
 we can write the Landau-level broadening as
\begin{equation}
\varGamma=(ev_{\rm F}/\pi)[2B\alpha/(N\mu)]^{1/2}.
\end{equation}

In the present study we consider the case of $n$-doping, i.e. the
Fermi level is high enough above the energy zero of the Dirac cone
in the range of ``+''-branch levels and the states of ``$-$''-branch
levels are completely filled, that they are irrelevant to electron
transport.

Special attention has to be paid to the $n=0$ level, since,
depending on the direction of exchange potential the effective
g-factor of a TI surface state,  $g_z$, can be positive, zero or
negative.\cite{ZGWang2010,Taskin2011} The sign and magnitude of the
effective g-factor determines how many states of the zero level
should be included in or excluded from the available states for
electron occupation in the case of $n$-doping at a magnetic field.
(i) If $g_z=0$, the $n=0$ level center is exactly at
$\varepsilon_0=0$ and the system is electron-hole symmetric. The
total number of negative energy states (including the states of the
lower half of the $n=0$ level and states of the ``$-$"-branch
levels) and that of positive energy states (including the states of
the upper half of the $n=0$ level and states of the ``$+$"-branch
levels) do not change when changing magnetic field. Therefore, the
lower-half negative energy states of this level are always filled
and the upper-half positive-energy states of it are available for
the occupation of particles which are counted as electrons
participating in transport in the case of $n$-doping. (ii) For a
finite positive $g_z>0$, the $n=0$ level $\varepsilon_0$ moves
downward to negative energy and its distance to the nearest
``$-$"-branch level is $2|\delta_z|=g_z\mu_{\rm B}B$ closer than to
the nearest ``+"-branch level at finite magnetic field strength $B$.
This is equivalent to the opening of an increasingly enlarged (with
increasing $B$) energy gap between the ``+"-branch states and the
states of the zero-level and the ``$-$"-branch levels. The opening
of a sufficient energy gap implies that with increasing magnetic
field the states in the ``+"-branch levels would no longer shrink
into the zero-level, and thus
 the $n=0$ level should be completely excluded from the conduction band,
 i.e. only particles occupying the ``+"-branch states
are counted as electrons participating in transport in the case of
$n$-doping, when the magnetic field $B$ gets larger than a certain
value (depending on the magnitude of $g_z$). (iii) For a finite
negative $g_z<0$, the $n=0$ level $\varepsilon_0$ moves upward to
positive energy and an increasingly enlarged energy gap will be
opened between the states of the zero-level and the ``+"-branch and
the states of ``$-$"-branch levels, and particles occupying the
$n=0$ level and ``+"-branch states are electrons participating in
transport when the magnetic field $B$ gets larger than a certain
value.

As a result, the experimentally accessible sheet density  $N$ of electrons participating in transport
is related to the Fermi energy $\varepsilon_{\rm F}$ by the following equation valid at finite $g_z$
for the magnetic field $B$ larger than a certain value:
\begin{equation}\label{}
N=-\frac{1}{2(\pi l_{\rm
B})^2} \int d\epsilon
f(\epsilon)\sum_n^{\infty} {\rm Im}G_n(\epsilon),
\end{equation}
in which $f(\epsilon)=\{\exp[(\epsilon-\varepsilon_{\rm F})/T]+1\}^{-1}$ is the Fermi distribution function
at temperature $T$ and the summation index $n$ goes over $(1,2,....)$ for $g_z>0$, or $(0,1,2,....)$ for $g_z<0$.
In the case of $g_z=0$,
\begin{equation}\label{}
N=-\frac{1}{2(\pi l_{\rm
B})^2}\int d\epsilon
f(\epsilon)\bigg[\sum_{n=1}^{\infty} {\rm Im}G_n(\epsilon)+ {\rm Im}G_0^{p}(\epsilon)\bigg]
\end{equation}
valid for arbitrary magnetic field, in which ${\rm Im}G_0^{p}(\epsilon)={\rm Im}G_0(\epsilon)\theta(\epsilon)$.

The imaginary part of relative-electron density correlation function
in the presence of a magnetic field, $\varPi_2(\bm q,\omega)$, can
be expressed in the Landau representation as\cite{Ting1977,cai1985}
\begin{equation}\label{piqw}
\varPi_2(\bm q,\omega)=\frac{1}{2\pi l_{\rm
B}^2}\sum_{n,n'}C_{n,n'}(l_{\rm B}^2q^2/2)\varPi_2(n,n',\omega),
\end{equation}
in which the transform factor
\begin{align}
C_{n,n'}(\xi)\equiv & \frac{e^{-\xi}\xi^{n_2-n_1}}{\mathcal R_n\mathcal R_{n'}}\frac{n_1!}{n_2!}\bigg[L_{n_1}^{n_2-n_1}(\xi)\nonumber\\
&\hspace{1.0cm}+s_ns_{n'}\mathcal P_n\mathcal
P_{n'}\sqrt{\frac{n_2}{n_1}}L_{n_1-1}^{n_2-n_1}(\xi)\bigg]^2,
\end{align}
 with
$n_1={\rm min}(n,n')$, $n_2={\rm max}(n,n')$, $s_n=1-\delta_{n,0}$,
and $L_n^m(x)$ being associated Laguerre polynomials. The
Landau-representation correlation function $\varPi_2(n,n',\omega)$
in Eq.\,(\ref{piqw}) can be constructed with the imaginary part of
the retarded Green's function ${\rm Im}G_{n}(\epsilon)$, or the
density-of-states, of the $n$th Landau level
as\cite{Ting1977,cai1985}
\begin{align}\label{p2nn}
\varPi_2(n,n',\omega)=&-\frac{1}{\pi}\int d\epsilon[f(\epsilon)-f(\epsilon+\omega)]\nonumber\\
&\hspace{1.2cm}\times{\rm
Im}G_n(\epsilon+\omega){\rm Im}G_{n'}(\epsilon).
\end{align}
The summation indices $n$ and $n'$ in Eq.\,(\ref{piqw}) are taken over $(1,2,....)$ for $g_z>0$, or $(0,1,2,...)$ for $g_z<0$.
In the case of $g_z=0$, Eq.\,(\ref{piqw}) still works and the summation indices $n$ and $n'$ go over $(0,1,2,...)$
but with ${\rm Im}G_{0}(\epsilon)$ replaced
by ${\rm Im}G_{0}^{p}(\epsilon)$ in Eq.\,(\ref{p2nn}).

\section{numerical results and discussions}
Numerical calculations are performed for the magnetoresistivity
$R_{xx}$ of surface state in a uniform TI Bi$_2$Se$_3$. At zero
temperature the elastic scattering contributing to the resistivity
is modeled by a Coulomb potential due to charged
impurities:\cite{Dimitrie,wang2011} $U(\bm
q)=n_ie^2/(2\epsilon_0\kappa q)$ with $n_i$ being the impurity
density, which is determined by the zero-magnetic-field mobility
$\mu$. At temperatures higher than $50\,{\rm K}$,\cite{Tang2011}
phonon scatterings play increasingly important role and the dominant
inelastic contribution comes from optical phonons. For this polar
material, the scattering by optical phonons via the deformation
potential can be neglected. Hence, we take account of inelastic
scattering from optical phonons via Fr\"{o}hlich coupling: $|M(\bm
Q)|^2=e^2{\it
\Omega}/(2\epsilon_0Q^2)(\kappa^{-1}_\infty-\kappa^{-1})$. In the
numerical calculation we use the following
parameters:\cite{CXLiu2010,madelung2004semiconductors,Dimitrie,Zhu186102}
Fermi velocity $v_{\rm F}=5.0\times 10^{5}\,{\rm m/s}$, static dielectric
constant $\kappa=100$, optical dielectric constant
$\kappa_\infty=20$, and phonon energy ${\varOmega}=7.4\,{\rm meV}$.
The broadening parameter is taken to be $\alpha=3$.

\begin{figure}
\begin{center}
  \includegraphics[width=0.4\textwidth]{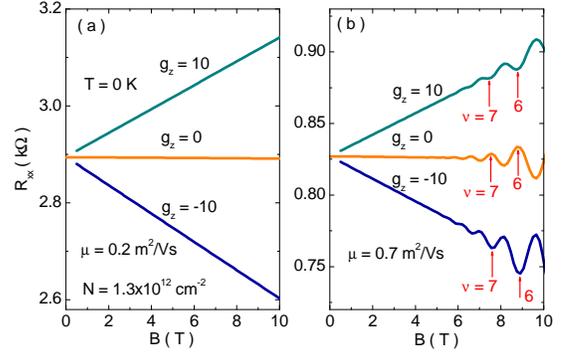}
\end{center}
\caption{(Color online) The calculated resistivity $R_{xx}$ as a
function of the magnetic field $B$ having different effective g-factors: $g_z=0, 10$ and $-10$
for a TI surface system with electron sheet density
$N=1.3\times10^{12}\,{\rm cm}^{-2}$ in the cases of
zero-magnetic-field mobility $\mu=0.2\,{\rm m^2/Vs}$ (a) and $\mu=0.7\,{\rm m^2/Vs}$ (b).
Several integer-number positions of filling factor $\nu=2\pi N/(eB)$ are marked in (b).}\label{diffg}
\end{figure}

Fig.\,\ref{diffg} shows the calculated magnetoresistivity $R_{xx}$ versus
the magnetic field strength $B$ for a TI surface system with electron sheet density
$N=1.3\times10^{12}\,{\rm cm}^{-2}$ but having different effective g-factors:
$g_z=0, 10$ and $-10$ for two values of zero-magnetic-field mobility
$\mu=0.2\,{\rm m^2/Vs}$ and $\mu=0.7\,{\rm m^2/Vs}$,
representing different degree of Landau-level broadening.
In the case without Zeeman splitting ($g_z=0$) the resistivity $R_{xx}$ exhibits
almost no change with changing magnetic field up to 10\,T, except the Shubnikov-de Haas (SdH)
oscillation showing up
in the case of $\mu=0.7\,{\rm m^2/Vs}$. This kind of magnetoresistance behavior
was indeed seen experimentally in the electron-hole symmetrical massless system of single-layer graphene.\cite{tan2011shubnikov}
In the case of a positive g-factor, $g_z=10$, the magnetoresistivity increases linearly with increasing magnetic field;
while for a negative g-factor, $g_z=-10$, the magnetoresistivity decreases linearly with increasing magnetic field.

\begin{figure}
\begin{center}
  \includegraphics[width=0.47\textwidth]{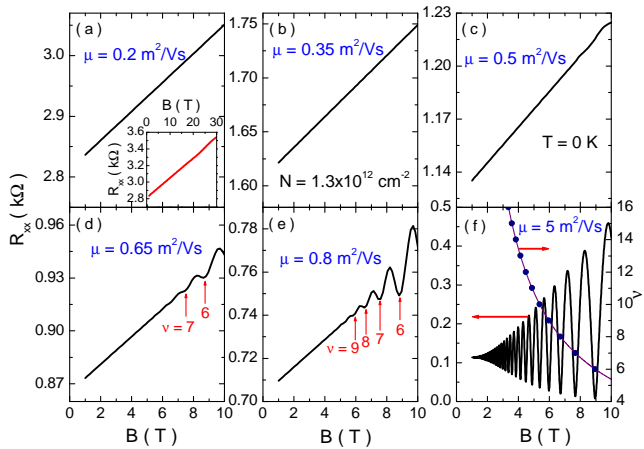}
\end{center}
\caption{(Color online) The longitudinal resistivity $R_{xx}$ is shown
as a function of the magnetic field $B$ for different values of
zero-magnetic-field mobility: (a) $\mu=0.2$, (b) $0.35$, (c) $0.5$,
(d) $0.65$, (e) $0.8$, and (f) $5\,{\rm m^2/Vs}$. The inset of (a)
illustrates the same for a larger magnetic-field range $0<B<30\,{\rm T}$.
The filling factor $\nu$ is plotted versus the magnetic field in (f);
and several integer-number positions of $\nu$ are also marked in (d) and (e).
Here the surface electron density $N=1.3\times10^{12}\,{\rm cm}^{-2}$ and the
lattice temperature $T=0\,{\rm K}$.}\label{rhoB}
\end{figure}

In the following we will give more detailed examination on the linearly increasing magnetoresistance
in the positive $g_z$ case.

Fig.\,\ref{rhoB} shows the calculated resistivity $R_{xx}$ versus
the magnetic field strength $B$ at lattice temperature $T=0\,{\rm K}$
for system of carrier sheet density
$N=1.3\times10^{12}\,{\rm cm}^{-2}$ and $g_z=10$, having different zero-field mobility
$\mu=0.2,0.35,0.5,0.65,0.8$ and $5\,{\rm m^2/Vs}$.
 All resistivity curves for mobility $\mu \leq 0.8\,{\rm
m^2/Vs}$ exhibit clear linearity in the magnetic-field range and
appear no tendency of saturation at the highest field shown in the figure.
Especially, for
the case $\mu=0.2\,{\rm m^2/Vs}$, the linear behavior extends even up
to the magnetic field of $30\,{\rm T}$, as illustrated in the inset of Fig.\,\ref{rhoB}(a).
This feature contradicts the classical MR which saturates at
sufficiently large magnetic field $B\gg\mu^{-1}$.

Note that here we only present the calculated $R_{xx}$ for magnetic
field $B$ larger than $B_c=1$\,T, for which a sufficient energy gap
$2|\delta_z|=g_z\mu_{\rm B}B$ is assumed to open that with further
increase of the magnetic field the states in the ``+''-branch levels
no longer shrink into the zero level and thus it should be excluded
from the conduction band. This is of course not true for very weak
magnetic field. When $B\rightarrow 0$ the energy gap
$2|\delta_z|\rightarrow 0$, the situation becomes similar to the
case of $g_z=0$: the whole upper half of the zero-level states are
available to electron occupation and we should have a flat
resistivity $R_{xx}$ when changing magnetic field. With increasing
$B$ the portion of the zero-level states available to conduction
electrons decreases until the magnetic field reaches $B_c$. As a
result the resistivity $R_{xx}$ should exhibit a crossover from a
flat changing at small $B$
to positively linear increasing at $B>B_c$. This is just the behavior
observed in the TI Bi$_2$Se$_3$.\cite{Tang2011}

Note that in the case of $\mu=0.2\,{\rm m^2/Vs}$, the broadened
Landau-level widths are always larger than the neighboring level
interval: $2{\it \Gamma}\gtrsim
\Delta\varepsilon_n=\varepsilon_{n+1}-\varepsilon_n$, which requires
$\mu\lesssim (4e\alpha/N)[(\sqrt{n+1}+\sqrt{n})/\pi]^2$, even for
the lowest Landau level $n=1$, i.e. the whole Landau-level spectrum
is smeared. With increasing the zero-field mobility the magnitude of
resistivity $R_{xx}$ decreases, and when the broadened Landau-level
width becomes smaller than the neighboring level interval, $2{\it
\Gamma}\lesssim \Delta\varepsilon_n$, a weak SdH oscillation begin
to occur around the linearly-dependent average value of $R_{xx}$ at
higher portion of the magnetic field range, as seen in
Fig.\,\ref{rhoB}\,(c), (d) and (e) for $\mu=0.5,0.65$ and $0.8\,{\rm
m^2/Vs}$. On the other hand, in the case of large mobility, e.g.
$\mu= 5\,{\rm m^2/Vs}$, where the broadened Landau-level widths
$2{\it \Gamma}$ are much smaller than the neighboring level interval
even for level index $n$ as large as $30$, the magnetoresistivity
shows pronounced SdH oscillation and the linear-dependent behavior
disappears, before the appearance of quantum Hall
effect,\cite{Zheng2002,zhang2005experimental,Gusynin2005} as shown
in Fig.\,\ref{rhoB}(f).

Abrikosov's model for the LMR requires the applied magnetic field large
enough to reach the quantum limit at which all the carriers are within
the lowest Landau level,\cite{Abrikosov1998} while it is obvious
that more than one Landau levels are occupied in the experimental samples
in the field range in which the linear and
non-saturating magnetoresistivity was observed.\cite{Tang2011} For the given electron surface density
$N=1.3\times 10^{12}\,{\rm cm}^{-2}$, the number of occupied Landau
levels, or the filling factor $\nu=2\pi N/(eB)$, at different magnetic fields
is shown in Fig.\,\ref{rhoB}(f), as well as in the Fig.\,\ref{rhoB}(d) and (e),
where the integer-number positions of ${\nu}$, i.e. filling up to entire $\nu$ Landau levels,
coincide with the minima of the density-of-states or the dips of SdH oscillation.
This is in contrast with $g_z=0$ case, where
the integer number of ${\nu}$, which implies a filling up to the center position of the $\nu$th Landau levels,
locates at a peak of SdH oscillation, as shown in Fig.\,\ref{diffg}b.
The observed SdH oscillations in the Bi$_2$Se$_3$ nanoribbon exhibiting nonsaturating
surface LMR in the experiment\cite{Tang2011}
favor the former case: a finite positive effective $g_z>0$.

\begin{figure}
\begin{center}
  \includegraphics[width=0.4\textwidth]{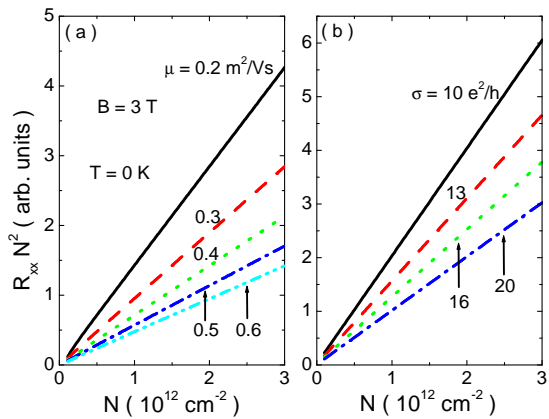}
\end{center}
\caption{(Color online) $R_{xx}N^2$ is plotted as a function of
the surface electron density $N$ at magnetic field $B=3\,{\rm
T}$: (a) at different values of zero-field mobility $\mu$,
and (b) at different values of zero-field conductivity $\sigma$.}\label{rhoN}
\end{figure}

\begin{figure}
\begin{center}
  \includegraphics[width=0.35\textwidth]{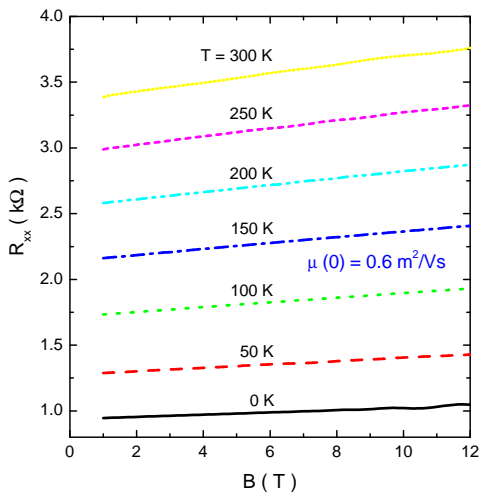}
\end{center}
\caption{(Color online) The longitudinal resistivity of the surface
state of a TI versus magnetic field $B$ at various lattice
temperatures. Here the zero-magnetic-field mobility at zero
temperature is $\mu(0)=0.6\,{\rm m^2/Vs}$.}\label{rhoT}
\end{figure}

Next, we examine the density-dependence of the linear
magnetoresistivity. To compare with Abrikosov's quantum
magnetoresistance which suggests a $R_{xx}\propto N^{-2}$
behavior,\cite{Abrikosov1998,Abrikosov2003} we show the calculated
$R_{xx}N^2$ for above LMR versus the carrier sheet density $N$ in
Fig.\,\ref{rhoN} at fixed magnetic field $B=3$\,T. The mobility is
taken respectively to be $\mu=0.2,0.3,0.4,0.5$ and $0.6$\,m$^2$/Vs
to make the resistivity in the LMR regime. A clearly linear
dependence of $R_{xx}N^{2}$ on the surface density $N$ is seen in
all cases, indicating that this non-saturating linear resistivity is
almost inversely proportional to the carrier density.  In the figure
we also show $R_{xx}N^2$ versus $N$ under the condition of different
given conductivity $\sigma=Ne\mu=10,13,16$ and $20\,e^2/h$. In this
case the half-width $\varGamma$ is independent of surface density.
The linear dependence still holds, indicating that this linear
behavior is  not sensitive to the modest $N$-dependence of Landau
level broadening $\varGamma$ as long as the system is in the
overlapped Landau level regime.

From the above discussion, it is obvious that LMR shows up in the
system having overlapped Landau levels and the separation of Landau
levels makes the MR departure from the linear increase. At high
temperature, the thermal energy would smear the level separation and
phonon scatterings further broaden Landau levels. Hence, it is
believed that this LMR will be robust against raising temperature.
This is indeed the case as seen in Fig.\,\ref{rhoT}, where we plot
the calculated magnetoresistivity $R_{xx}$ for the above system with
zero-temperature linear mobility $\mu(0)=0.6$\,m$^2$/Vs versus the
magnetic field at different lattice temperatures. We can see that
raising temperature to room temperature has little effect on the
linearity of MR. Due to the decreased mobility at higher temperature
from phonon scattering, the weak SdH oscillation on the linear
background tends to vanish. These features are in good agreement
with the experimental report.\cite{Tang2011}

\section{summary}
In summary, we have studied the two-dimensional magnetotransport in the flat
surface of a three-dimensional TI, which arises from the surface states with a wavevector-linear energy
dispersion and a finite, positive Zeeman splitting within the bulk energy gap.
When the level broadening is comparable to or larger
than the Landau-level separation and the conduction electrons spread
over many Landau levels, a positive, dominantly linear and
non-saturating magnetoresistance appears within a quite wide range
of magnetic field and persists up to room temperature. This
remarkable LMR provides a possible mechanism for the recently
observed linear magnetoresistance in topological insulator
Bi$_2$Se$_3$ nanoribbons.\cite{Tang2011}

In contrast to quantum Hall effect which appears in the case of well formed Landau levels
and to Abrikosov's quantum magnetotransport,\cite{Abrikosov1998,abrikosov2000quantum}
 which is limited to the extreme
quantum limit that all electrons coalesce into the lowest Landau
level, the discussed LMR is a phenomena of pure classical
two-dimensional magnetotransport in a system having
linear-energy-dispersion, appearing in the regime of overlapped
Landau levels, irrespective of its showing up in relatively high
magnetic field range. Furthermore, the present scheme deals with
spatially uniform case without invoking the mobility fluctuation in
a strongly inhomogeneous system, which is required in the classical
Parish and Littlewood model to produce a LMR.\cite{Parish2003}

The appearance of this significant positive-increasing linear
magnetoresistance depends on the existence of a positive and sizable
effective g-factor. If the Zeeman energy splitting is quite small
the resistivity $R_{xx}$ would exhibit little change with changing
magnetic field. In the case of a negative and sizable effective
g-factor the magnetoresistivity would decrease linearly with
increasing magnetic field. Therefore, the behavior of the
longitudinal resistivity versus magnetic field may provide a useful
way for judging the direction and the size of the effective Zeeman
energy splitting in TI surface states.

\section*{ACKNOWLEDGMENTS}
This work was supported by the National Science Foundation of China
(Grant No. 11104002), the National Basic Research Program of China (Grant No. 2012CB927403)
and by the Program for Science\&Technology Innovation
Talents in Universities of Henan Province (Grant No. 2012HASTIT029).


\begin{thebibliography}{10}

\bibitem{xu1997large}
R.~Xu, A.~Husmann, T.~F. Rosenbaum, M.~L. Saboungi, J.~E. Enderby,
and P.~B. Littlewood, Nature \textbf{390}, 57 (1997).

\bibitem{hu2008classical}
J. Hu and T. Rosenbaum, Nature Mater. \textbf{7}, 697 (2008).

\bibitem{delmo2009large}
M.~P. Delmo, S. Yamamoto, S. Kasai, T. Ono, and K. Kobayashi, Nature
  \textbf{457}, 1112 (2009).

\bibitem{Johnson2010}
H.~G. Johnson, S.~P. Bennett, R. Barua, L.~H. Lewis, and D. Heiman,
Phys. Rev. B \textbf{82}, 085202 (2010).

\bibitem{friedman2010quantum}
A.~L. Friedman, J.~L. Tedesco, P.~M. Campbell, J.~C. Culbertson,
E. Aifer,  F.~K. Perkins, R.~L. Myers-Ward, J.~K. Hite, C.~R. Eddy~Jr, G.~G. Jernigan,
  and D.~K. Gaskill, Nano Lett. \textbf{10}, 3962 (2010).

\bibitem{Kapitza1929}
P.~L. Kapitza, Proc. R. Soc. A \textbf{123}, 292 (1929).

\bibitem{Abrikosov1998}
A.~A. Abrikosov, Phys. Rev. B \textbf{58}, 2788 (1998).

\bibitem{abrikosov2000quantum}
A.~A. Abrikosov, Europhys. Lett. \textbf{49}, 789 (2000).

\bibitem{Parish2003}
M.~M. Parish and P.~B. Littlewood, Nature \textbf{426}, 162 (2003).

\bibitem{Kane2005}
C.~L. Kane and E.~J. Mele, Phys. Rev. Lett. \textbf{95}, 226801
(2005).

\bibitem{hasan2010colloquium}
M.~Z. Hasan and C.~L. Kane, Rev. Mod. Phys. \textbf{82}, 3045 (2010).

\bibitem{qi2010topological}
X.~L. Qi and S.~C. Zhang, Rev. Mod. Phys. \textbf{83}, 1057 (2011).

\bibitem{xia2009observation}
Y. Xia, D. Qian, D. Hsieh, L. Wray, A. Pal, H. Lin, A. Bansil,
D.~Grauer, Y.~S. Hor, R.~J. Cava, and M.~Z. Hasan, Nat. Phys. \textbf{5}, 398 (2009).

\bibitem{zhang2009natphys}
H.~J. Zhang, C.~X. Liu, X.~L. Qi, X.~Dai, Z.~Fang and S.~C. Zhang, Nat. Phys. \textbf{5}, 438 (2009).

\bibitem{CXLiu2010}
C.~X. Liu, X.~L. Qi, H.~J. Zhang, X. Dai, Z. Fang, and S.~C. Zhang, Phys. Rev. B \textbf{82},
045122 (2010).

\bibitem{Tang2011}
H.~Tang, D.~Liang, R.~L.~J. Qiu, and X.~P.~A. Gao, ACS Nano
\textbf{5}, 7510 (2011).



\bibitem{lei1985gsf}
X.~L. Lei and C.~S. Ting, Phys. Rev. B \textbf{30}, 4809 (1984);  Phys. Rev. B \textbf{32}, 1112 (1985).

\bibitem{lei1985tdb}
X.~L. Lei, J.~L. Birman, and C.~S. Ting, J. Appl. Phys. \textbf{58}, 2270 (1985).

\bibitem{cai1985}
W. Cai, X.~L. Lei, and C.~S. Ting, Phys. Rev. B \textbf{31}, 4070 (1985);
X.~L. Lei, W. Cai, and C.~S. Ting, J. Phys. C: Solid State, \textbf{18}, 4315 (1985).

\bibitem{Ting1977}
C.~S. Ting, S.~C. Ying, and J.~J. Quinn, Phys. Rev. B \textbf{16},
5394 (1977).

\bibitem{McClure1956}
J.~W. McClure, Phys. Rev. \textbf{104}, 666 (1956).

\bibitem{Zheng2002}
Y. Zheng and T. Ando, Phys. Rev. B \textbf{65}, 245420 (2002).

\bibitem{Zarea2005}
M. Zarea and S.~E. Ulloa, Phys. Rev. B \textbf{72}, 085342 (2005).

\bibitem{ZGWang2010}
Z.~G. Wang, Z.~G. Fu, S.~X. Wang, and P. Zhang, Phys. Rev. B
\textbf{82}, 085429 (2010).

\bibitem{Taskin2011}
A.~A. Taskin and Y. Ando, Phys. Rev. B \textbf{84}, 035301 (2011).

\bibitem{Ando1982}
T. Ando, A.~B. Fowler, and F. Stern, Rev. Mod. Phys. \textbf{54},
437 (1982).

\bibitem{Lei2003}
X.~L. Lei and S.~Y. Liu, Phys. Rev. Lett. \textbf{91}, 226805 (2003).

\bibitem{wang2011}
C.~M. Wang and F.~J. Yu, Phys. Rev. B \textbf{84}, 155440 (2011).

\bibitem{Dimitrie}
D. Culcer, E.~H. Hwang, T.~D. Stanescu, and S. Das~Sarma, Phys. Rev.
B  \textbf{82}, 155457 (2010).

\bibitem{madelung2004semiconductors}
O.~Madelung, {\it Semiconductors: Data Handbook} (Springer Verlag, 2004).

\bibitem{Zhu186102}
X. Zhu, L. Santos, R. Sankar, S. Chikara, C.~. Howard, F.~C. Chou,
C. Chamon,  and M. El-Batanouny, Phys. Rev. Lett. \textbf{107}, 186102 (2011).

\bibitem{tan2011shubnikov}
Z.~B. Tan, C.~L. Tan, L. Ma, G.~T. Liu, L. Lu, and C.~L. Yang, Phys. Rev.
B \textbf{84}, 115429 (2011).

\bibitem{zhang2005experimental}
Y. Zhang, Y. Tan, H. Stormer, and P. Kim, Nature \textbf{438}, 201 (2005).

\bibitem{Gusynin2005}
V.~P. Gusynin and S.~G. Sharapov, Phys. Rev. Lett. \textbf{95},
146801 (2005).


\bibitem{Abrikosov2003}
A.~A. Abrikosov, J. Phys. A: Math. Gen. \textbf{36}, 9119 (2003).


\end{thebibliography}
\end{document}